\documentstyle[12pt]{article}

\newcommand{\partud}{
{d \sigma (\uparrow\uparrow-\uparrow\downarrow) \over dx dCos(\theta)}}
\newcommand{\partc}{
{d \hat \sigma^{a(h)} (s_1) \over dy dCos(\theta)}}
\newcommand{\partD}{
{d \Delta \hat \sigma^a  \over dx dCos(\theta)}}
\newcommand{\partDD}{
{d \Delta \hat \sigma^a  \over dy dCos(\theta)}}
\newcommand{\parto}{
{d \sigma^0 (\uparrow\uparrow-\uparrow\downarrow) 
\over dz dCos(\theta)}}
\newcommand{\partu}{
{d \hat \sigma^{a(\uparrow)}(\uparrow)  \over dx dCos(\theta)}}
\newcommand{\partd}{
{d\hat  \sigma^{a(\downarrow)}(\uparrow)  \over dx dCos(\theta)}}
\newcommand{\partq}{
{d \sigma^q  \over dx_1 dCos(\theta)}}
\newcommand{\partg}{
{d \sigma^g  \over dx_3 dCos(\theta)}}
\newcommand{\partv}{
{d \sigma^v (\uparrow\uparrow-\uparrow\downarrow) \over dx_1 dCos(\theta)}}
\newcommand{\partQ}{
{d \sigma^q (\uparrow\uparrow-\uparrow\downarrow) \over dx_1 dCos(\theta)}}
\newcommand{\partG}{
{d \sigma^g (\uparrow\uparrow-\uparrow\downarrow) \over dx_3 dCos(\theta)}}
\newcommand{\partss}{
{d \sigma (s,s_1) 
\over dx dCos(\theta)}}
\newcommand{\be}{\begin{equation}}
\newcommand{\ee}{\end{equation}}
\newcommand{\bea}{\begin{eqnarray}}
\newcommand{\eea}{\end{eqnarray}}
\begin{document}
\begin{titlepage}
\vspace*{\fill}
\begin{center}
{\Large \bf Order $\alpha_s(Q^2)$ QCD corrections to \\
the polarised $e^+~e^- \rightarrow \Lambda ~X$}\\[1cm]
{\bf V. Ravindran}\\
{\em Theory Group, Physical Research Laboratory, Navrangpura \\
Ahmedabad 380 009, India}\\
\end{center}
\vspace{2cm}
\begin{abstract}
The importance of polarised gluons fragmenting into
$\Lambda$ in the polarised $e^+ e^-$ annihilation
is discussed using the Altarelli-Parisi evolution equations
satisfied by the quark and gluon fragmentation
functions.  In this context, the polarised fragmentaion function
$\hat g_1^\Lambda(x,Q^2)$ appearing in the cross section
is discussed within the parton model.  We relate 
this fragmentation function to the quark, anti-quark 
and gluon fragmentation functions and also find the QCD 
corrections to order $\alpha_s$.  

\end{abstract}
\vspace*{\fill}
\end{titlepage}
Quantum Chromodynamics(QCD) has been the most successful theory of 
strong interaction physics.  A number of Deep Inelastic Scattering(DIS) 
experiments not only tested the excellent predictions
of QCD such as scaling violation but also unravelled the structure
of the hadrons in terms of its fundamental constituents 
such as quarks and gluons.  The careful study of the unpolarised 
lepton proton DIS showed that the gluons carry significant 
amount of momentum of the proton.  This is an interesting result 
as it contradicts our naive expectation that the gluonic 
contribution is next to leading order effect.  This naive 
expectation is due to simple observation that 
one has to pay price for the strong coupling constant when photon
probes the gluonic content of the proton.  This simple argument 
based on strong coupling constant fails due to the appearance 
of large logarithms at very high energies.  More sophisticated 
analysis based on the Altarelli-Parisi(AP) evolution equations 
satisfied by the parton probability distribution functions 
confirms the importance of the role played by the
gluons $\cite {MUTA}$.  

More recently, a series of DIS 
experiments with polarised beam and
polarised target helped us to understand the spin structure of the proton.
These experiments measured the polarised structure function 
$g_1(x,Q^2)$ of the proton.  Here $x$ is the usual Bjorken variable
and $Q^2$ is the invariant mass of the probing photon.
The first moment of this structure function directly measures 
the spin contribution coming from the constituents of the polarised
proton.  To lowest order,
\be
\int_0^1 g_1(x,Q^2) dx = {1\over 2}\sum_q e_q^2 \left[\Delta q(Q^2)
+ \Delta \bar q(Q^2)\right]
\label{g1s}
\ee
where $e_q$ is the charge of the quark $q$ and $\Delta q(Q^2)=
\int_0^1\Delta q(x,Q^2) dx$.  Using the low energy data on 
$\Delta u(Q^2) - \Delta d(Q^2)$ and 
$\Delta u(Q^2)+\Delta d(Q^2) - 2 \Delta s(Q^2)$,
the measured first moment predicts that the spin contribution 
coming from the quarks is small.  The naive argument 
based on the fact that the gluons 
are next to leading order again fails and in fact the gluonic
contribution to the first moment of $g_1(x,Q^2)$ is found to be large.
The reason for this is that the AP evolution equation for the polarised gluon
distribution function shows that $\alpha_s(Q^2) \Delta g(Q^2)$ is 
scale independent to order $\alpha_s(Q^2)$(strong coupling constant, 
$\alpha_s(Q^2) = g_s^2/4 \pi$) $\cite {ANS}$.  
Crudely speaking, large logarithms appearing in the high 
energy scattering cross sections and the running of strong coupling constant
invalidate our naive expectations based on simple counting rules. 

Hadron production in $e^+ e^-$ annihilation
plays the crucial role as DIS does to understand the structure of the hadron. 
DIS experiments help us to understand how quarks and gluons share the
properties of hadron such as its momentum and its spin.  On the other hand,
hadroproduction in $e^+ e^-$ annihilation is useful to 
understand the fragmentation mechanism of quarks and gluons into hadrons
$\cite {FEYN}$.  
There has been a number of experiments
on unpolarised $e^+ e^-$ experiments and a lot of theoretical
works to understand the data $\cite {NASON}$.  All these 
analysis show that QCD
corrections are very important.  More recently, a series of experiments 
have been done at CERN measuring polarisation of 
hyperons produced $\cite{LUN}$.  On the theoretical 
side,  
there has been a lot of interesting developments.  
Burkardt and Jaffe $\cite {BUR}$ have proposed an experimental
programme to successfully measure the polarised fragmentation functions.
This programme requires the measurement of the total inclusive $\Lambda$
production in polarised $e^+ e^-$ annihilation at various energies.

Recently a systematic analysis $\cite {RAV}$ of the evolution of polarised 
quark and gluon fragmentation functions has been done using
the AP evolution equations satisfied by these polarised fragmentation 
functions.  The analysis showed that the polarised
gluon fragmentation is as significant as quark fragmentation functions.
So in this letter we calculate the QCD corrections to 
polarised $e^+ e^-$
scattering to order $\alpha_s(Q^2)$ in the parton model.  This consists
of two sectors. 1.  Quark sector and 2. Gluon sector.  In the quark sector,
the quark and/or anti quark is polarised.   They are QCD corrected by
gluon bremstalungs in addition to virtual corrections
due to gluons.  This changes the probability of a polarised
quark and/or antiquark fragmenting into a polarised hadron.  
In the gluon sector, 
the contribution starts at order $\alpha_s(Q^2)$.  This in fact measures
the probability of polarised gluon fragmenting into
polarised hadron.  All these processes to order $\alpha_s(Q^2)$
are singular in both soft and collinear limit when the masses
of quarks are taken to be zero.  We regulate them by giving 
a small mass $m_g$ to gluons.  Hence, the polarised fragmentation 
functions appearing in the cross section formula are defined in this scheme.  
We work in the "massive gluon scheme" as it
regulates both soft and collinear divergences simultaneously.  
The other reason why we work in the massive gluon scheme 
is that in the polarised DIS experiments, the data can be better 
understood in terms of polarised gluon distributions defined 
in the massive gluon scheme.  It is worth recalling that 
the total cross section measured in the laboratory
is nothing to do with the scheme we choose.  Hence, the choice
is immaterial as for as the physical predictions are concerned.  
In the quark sector we encounter Ultraviolet(UV) divergences too.  We regulate 
them in the well known gauge invariant Pauli-Villar's regularisation. 
All these divergences are shown to cancel 
among themselves, thanks to Ward Identity.  Since we set quark masses
to be zero, the convenient basis is helicity basis.  

Let us first consider 
the AP evolution equation satisfied by the polarised quark and gluon 
fragmentation functions.
\be
{d\over dt}{\pmatrix {\Delta D_{q_i}^H(x,t)\cr \Delta D_g^H(x,t)\cr}}
 = {\alpha_s(t) \over 2 \pi}\int_x^1 {dy \over y}
   {\pmatrix {\Delta P_{qq}(x/y) & 
   \Delta P_{gq}(x/y)\cr \sum_q \Delta P_{qg}(x/y) & 
   \Delta P_{gg}(x/y)\cr}} {\pmatrix {\Delta D_{q_i}^H(y,t)\cr 
   \Delta D_g^H(y,t)\cr}}
\label{matrix}
\ee
where $\Delta P_{ij}(x)$ are given in $\cite {MUTA}$ and 
$t=\log(Q^2 /\Lambda^2)$. 
The matrix $\Delta P$ is just the transpose
of what is appearing in the AP equation for polarised parton distribution
functions.  Noting that the first moments of $\Delta P_{qq}(z)$ and
$\Delta P_{qg}(z)$ are zero and that of $\Delta P_{gq}(z)$ and
$\Delta P_{gg}(z)$ are $2$ and $ 2 \pi \beta_0$ 
(where $\beta_0= (11 C_2(G) - 4 T(R) )/12 \pi$ with
$C_2(R)= 4/3$, $C_2(G)=3$ and $T(R)=f/2$, $f$ being the
number of flavours) respectively,
we find that the first moment of polarised quark and gluon 
fragmentation functions
satisfy the following simple differential equations.
\bea
{ d \over dt} \Delta D_g^H(t) &=& \alpha_s(t) \beta_0 \Delta D_g^H(t) \\
{ d \over dt} \Delta D_q^H(t) &=&{1 \over \pi} \alpha_s(t) \Delta D_g^H(t)
\label{qgde}
\eea
The solution to these equations can be found very easily.  It 
turns out the polarised gluon fragmentation function satisfies
\be
{d \over dt} (\alpha_s(t) \Delta D_g^H(t))= 0 (\alpha_s(t)^3)
\label{paper}
\ee
where renormalisation group equation for $\alpha_s(t)$ i,e $d \alpha_s(t) /dt = -\beta_0 \alpha_s(t)^2$ 
has been used.  This is analogous to the differential equation satisfied by the
first moment of polarised gluon distribution function, 
i.e ${d \alpha_s(t) \Delta g(t) / dt} =0 $  to order $\alpha_s(t)$ 
where $\Delta g(t)$ is the first moment of
polarised gluon distribution function $\cite {ANS}$.  From eqn.(\ref{paper}),
we find that $\alpha_s(t) \Delta D_g^H(t)$ is constant.  
This implies that the first moment of polarised gluon 
fragmentation function grows logarithmically.
Using this, one finds that $\Delta D_q^H(t)$ is proportional to $t$.  
This analysis implies that
when one wants to understand how much of the spin is transferred 
to hadron, one must consider the direct gluon contribution also.
It is in this spirit we compute the gluonic contribution to polarised
$e^+ e^- $ scattering.  In particular, we consider 
the inclusive polarised $\Lambda$ 
hyperon production in the annihilation of polarised $e^+ e^-$ process.
The reason why we consider this is two fold.  Firstly the inclusive
$\Lambda$ production cross section is large compared to other particles 
and secondly it is easy to measure the polarisation of $\Lambda$.  
Our analysis can be extended to other
hadro productions.  We work in the energy region
where only photon channel is dominant.  The complete analysis 
including $Z$ exchange channel is reserved for future publication 
$\cite {RAV1}$.

We consider the inclusive polarised 
$\Lambda$ production rate which factorises as:
\be
d \sigma(s,s_1) = {1 \over 4 q_1.q_2} L^{\mu \nu} (q_1,q_2,s_1)
\left( {e^2 \over Q^4}\right)
4 \pi W_{\mu \nu}^\Lambda (q,p,s) {d^3 p \over (2 \pi)^3 2 p_0}
\label{fact}
\ee
where $L_{\mu \nu}(q_1,q_2,s_1)$ is the leptonic part arising from $e^+ e^-$
annihilation into a photon of virtuality $Q^2$ and
$W_{\mu \nu}^\Lambda(p,q,s)$ is photon fragmentation tensor.
The first two arguments of these tensors are momenta
described in the figures 1,2,3 and $s,s_1$ are the spins of the $\Lambda$
and the electron respectively.  The photon fragmentation tensor, sometimes
called hadronic tensor contains all the informations about 
the polarised quark and gluon fragmenting into hyperons.  

The operator definition of $W_{\mu \nu}^\Lambda(q,p,s)$ is found to be
\be
W_{\mu \nu}^\Lambda (q,p,s)= {1 \over 4 \pi} \int d^4 \xi e^{i q.\xi}
<0\vert J_\mu(0) \vert \Lambda(p,s) X><\Lambda(p,s) X\vert J_\nu(\xi)\vert 0>
\label{hadten}
\ee
where $J_\mu(\xi)$ is the electromagnetic($em$) current, 
$X$ is unobserved hadrons (summation over $X$ is implicit),
$p$ and $s$ are the momentum and spin of the $\Lambda$ detected.
Since we are interested in the polarised cross section and the
energy is not too high to produce $Z$ vector boson, only antisymmetric part of
leptonic and photonic tensors contribute to this cross section.
The antisymmetric parts of the leptonic tensor is found to be
\be
L_{\mu \nu}(q_1,q_2,s_1)= -2 i e^2 s_1 \epsilon_{\mu\nu\alpha\beta} q_1^\alpha 
q_2^\beta
\label{lepten}
\ee
where $q_1,q_2$ are the momenta of the incoming leptons and
$s_1$ is the spin of the polarised lepton(electron).
On the other hand the photonic tensor is not calculable
in Perturbative QCD(PQCD) as we do not know how to compute 
the matrix element of $em$ current 
between hadronic states and the vacuum.  But this can be
parametrised using Lorentz covariance, gauge invariance, Hermiticity,
and parity invariance.  The anti symmetric part of this 
tensor takes the following form $\cite {LU}$:
\be
W_{\mu \nu}^\Lambda(q,p,s)= {i\over p.q} \epsilon_{\mu \nu \lambda \sigma} 
q^\lambda s^\sigma \hat g_1^\Lambda(x,Q^2) + {i\over p.q} \epsilon_{\mu \nu 
\lambda \sigma} q^\lambda \left (s^\sigma- {s.q \over p.q} 
p^\sigma \right) \hat g_2^\Lambda(x,Q^2)
\label{hadexp}
\ee
where $x=2 p.q/Q^2$, $Q^2=q^2$ and $s^2=-1$.
Here the polarised fragmentation functions $\hat g_i^\Lambda(x,Q^2)$ 
are real and Lorentz 
invariant, hence they are functions of $x$ and $Q^2$.  
We have put hat on them to distinguish them from the 
polarised structure functions appearing in the polarised
DIS.  The following asymmetric cross section projects out
only the $\hat g_1^\Lambda(x,Q^2)$ structure function as  
\be
\partud=\alpha^2 {\pi \over Q^2} x \hat g_1^\Lambda(x,Q^2) Cos(\theta)
\label{asym}
\ee   
where $\alpha= e^2/4 \pi$, $\theta$ is the angle between 
the produced $\Lambda$ particle and the incoming electron. 
Here, $\uparrow \uparrow$ means that both incoming electron
and the produced hardron are paralelly polarised and $\uparrow \downarrow$
means that they are polarised antiparalelly. 
Note that $\hat g_2^\Lambda(x,Q^2)$ does not contribute to 
this asymmetry.  The above cross section is zero when 
we integrate over $\theta$.  This implies that
the polarised fragmentation function $\hat g_1^\Lambda(x,Q^2)$ 
can be measured only 
through angular distribution of the polarised hyperon.  
In the following, we interprete this structure function in the parton
model in terms of polarised quark and gluon fragmentation functions 
$\cite {FEYN}$.  

In the parton model the $\Lambda$ production inclusive cross section can 
be expressed in terms of inclusive quark and gluon production cross sections
convoluted with appropriate quark and gluon fragmentation functions.
\be
\partss=\sum_{a,h} \int_x^1 {dy \over y} \partc D_{a(h)}^{\Lambda(s)}(x/y,Q^2)
\label{parmod}
\ee
where $a$ runs over all the partons such as quarks, antiquarks and gluons 
and $h$ is their helicity.   
Here the left hand side is the parton differential cross section for the
production of parton of type $a$ with polarisation $h$ and energy fraction
$y=2 p.q/Q^2$ produced at an angle $\theta$ with 
respect to the beam direction(electron).
$D_{a(h)}^{\Lambda(s)}(z,Q^2)$ is the probability of a parton of type $a$ 
with polarisation $h$ fragmenting into $\Lambda$ with polarisation 
$s$ and the momentum fraction $z$ of the parent parton.  
In the parton model, the asymmetry we consider in eqn.(\ref{asym}) 
turns out to be
\be
\partud= \sum_a \int_x^1 {dy \over y} \partDD \Delta D_a^\Lambda(x/y,Q^2)
\label{parmodasy}
\ee
where 
\bea
\partD&=&\partu-\partd \nonumber\\
\Delta D_a^\Lambda(z,Q^2)&=&D_{a(\uparrow)}^{\Lambda(\uparrow)} (z,Q^2)
-D_{a(\downarrow)}^{\Lambda(\uparrow)}(z,Q^2)
\nonumber
\label{defns}
\eea
To arrive at this simple form, we have used parity 
invariance of the fragmentation functions.

In general $\hat g_1^\Lambda(x,Q^2)$ can be decomposed 
into the following parts:
\be
\hat g_1^\Lambda(x,Q^2)=\hat g_1^0(x,Q^2)+\hat g_1^q(x,Q^2)+\hat g_1^g(x,Q^2)
\label{genexp}
\ee
The first term is the lowest order contribution coming form Fig. 1.  
The second term $\hat g_1^q(x,Q^2)$ is $ \alpha_s(Q^2)$ correction coming 
from gluon bremstalung and virtual corrections to the polarised quark and
anti quark fragmenting into a polarised $\Lambda$.  The third term comes 
from the polarised gluon fragmenting into a polarised $\Lambda$.  
The lowest order term $\hat g_1^0(x,Q^2)$ can be computed from the Fig. 1. 
The asymmetry to lowest order is found to be
\be
\parto=3 \alpha^2 e_q^2 {\pi \over Q^2} \delta(1-z) Cos(\theta)
\label{asy0}
\ee
where $e_q$ is the charge of the quark. 
This is convoluted with appropriate quark and antiquark polarised 
fragmentation functions
to get $\hat g_1^0(x,Q^2)$(see parton model expression eqn.(\ref{parmodasy})):
\be
\hat g_1^0(x,Q^2)=3{1\over x}\sum_q e_q^2 \left( \Delta D_q^\Lambda(x,Q^2) 
+ \Delta D_{\bar q}^\Lambda(x,Q^2)
\right)
\label{g10}
\ee

Next we compute $\hat g_1^q(x,Q^2)$.  We first compute the real gluon emission
i.e bremstalung contribution to it.  These processes are both 
soft and collinear singular when we keep all the masses zero.
In order to regulate these two divergences simultaneously
we give a small mass $m_g$ to the gluons.  The gluon bremstalung contribution
to this process can be formally expressed as
\be
\partq={i s \over 4 q_1.q_2} L_{\mu \nu} (q_1,q_2,s_1) {1 \over Q^4}
\epsilon_{\mu\nu\lambda\sigma} q^\lambda p_1^\sigma {Q \over p_1.q} H_q(p_1,q)
\label{asyq1}
\ee
where $x_1=2p_1.q/Q^2$,
$p_1$ is the momentum of the polarised anti-quark, $q=q_1+q_2$, 
$\theta$ is the angle between produced anti-quark and the incoming electron 
and $s$ is its spin.  The projected hard part of the 
Fig. 2, $H_q(p_1,q)$ is given by
\be
H_q(p_1,q)= {Q \over 32 (2 \pi)^3} \int dx_2 {\cal P}_q^{\mu \nu} \vert M^q 
\vert^{2}_{\mu \nu}
\label{hadq}
\ee
where $x_2=2p_2.q/Q^2$ and the projector 
${\cal P}_q=i \epsilon_{\mu \nu \lambda \sigma} p_1^\lambda q^\sigma/2 p_1.q$.  
In terms of Mandelstam variables, the projected matrix element square
${\cal P}_q. \vert M^q\vert^2$ is found to be
\bea
{\cal P}_q^{\mu \nu} \vert M^q \vert^{2}_{\mu \nu}&=&{32 (2 \pi)^3 
e_q^2 \alpha \alpha_s \over  \pi} 
\left [ 2 {s+t-m_g^2-Q^2 \over s t (Q^2-t)} (s t +Q^2 t -Q^4 
-m_g^2 Q^2) \right. \nonumber \\
&& \left. +{s t -m_g^2 Q^2 \over t^2} + {s t - m_g^2 Q^2 \over s^2 
(t-Q^2)}(t + 2 s -2 m_g^2 -Q^2) \right]
\label{prhadq}
\eea
where the Mandelstam variables $s=(p_1+p_3)^2$,$t=(p_2+p_3)^2$ and 
$u=(p_1+p_2)^2$ satisfy $s+t+u=m_g^2+Q^2$.
Noting that $s=Q^2(1-x_2)$ and $t=Q^2(1-x_1)$ and 
using eqns.(\ref{asyq1},\ref{hadq},\ref{prhadq}), the differential cross 
section for the emission of real
gluons is found to be
\bea
\partq&=& { 2 e_q^2 \alpha^2 \alpha_s \over Q^2} \left[
\left( {1+x_1^2 \over 1-x_1} \right) \log\left({Q^2 x_1 
(1-x_1) \over m_g^2}\right) -{3 \over 2}(1-x_1) \right. \nonumber\\
&&\left.- {3 \over 2} {1 \over (1-x_1)} + {5 \over4} \delta(1-x_1) 
\right] Cos(\theta)
\label{asyq2}
\eea
Note that there are two types of singularities appearing in the
expression when we take $x_1\rightarrow 0$ and $m_g \rightarrow 0$.

Next we compute the virtual gluon contribution to this differential
cross section.  It involves the evaluation of self energy 
and vertex corrections (see Fig. 3).  Both diagrams in Fig. 3
are UV divergent separately.
As we have already mentioned, we regulate these UV divergences 
in Pauli-Villar's regularisation scheme.  They also suffer 
from soft divergences when we set all the masses to
be zero.  Here also, we give a small mass to the gluon to regulate
them.  The amplitude for the self energy insertion
after properly taking care of wave function renormalisation
turns out to be(self energies for both quark and antiquark)
\be
M_\mu^{self}= {i \over  \pi} e_q e \alpha_s \log\left({L \over m_g^2}\right)
\bar v_{s}(p_1) \gamma_\mu u_{s'}(p_2)
\label{self}
\ee
where  $L$ is the Pauli-Villar's regulator to regulate the UV divergence.
$s'$ and $s$ are the polarisations of quark and antiquark respectively.

Similarly the amplitude for the vertex correction turns out to be
\bea
M_\mu^{vertex}&=&{i \over  \pi} e_q e \alpha_s \left[\log^2\left({m_g^2 \over Q^2}\right)
+3 \log\left({m_g^2 \over Q^2}\right)\right. \nonumber \\
&& \left.+ {7 \over 2} -{\pi^2  \over 3}
-\log\left({L \over m_g^2}\right)\right] \bar v_{s}(p_1) 
\gamma_\mu u_{s'}(p_2)
\label{vert}
\eea
Though the above amplitudes suffer from UV divergence,
when you sum these two amplitudes, we find that
the UV regulator cancels between self energy and vertex corrections, 
thanks to Ward identity.  Hence the virtual contribution 
to the partial differential cross section is found to be
\bea
\partv&=&{2 e_q^2 \alpha^2 \alpha_s \over Q^2} 
\left[-\log^2\left({m_g^2 \over Q^2}\right)
-3 \log\left({m_g^2 \over Q^2}\right)\right. \nonumber \\
&& \left. -{7 \over 2} + {\pi^2 \over 3}\right] \delta(1-x_1)
Cos(\theta)
\label{vir}
\eea

Adding the virtual and bremstalung gluon contributions we obtain
\bea
\partQ &=&{2 e_q^2 \alpha^2 \alpha_s \over Q^2} \left [
\left({1+x_1^2 \over 1-x_1}\right)_+ \log \left({Q^2 \over m_g^2}\right) 
- {3 \over 2} (1-x_1) 
\right. \nonumber \\
&& \left. +(1+x_1^2) \left ({\log(1-x_1) \over 1-x_1}\right)_+ 
+{1+x_1^2 \over 1-x_1} \log(x_1) 
\right. \nonumber \\
&&\left. - {3 \over 2} \left ({1\over 1-x_1}\right )_+ 
-\left ( {9 \over 4} - {\pi^2 \over 3} \right) \delta(1-x_1) \right]
Cos(\theta)
\label{asyq3}
\eea
where the $+$ prescription is defined in the usual way.  That is,
\be
\int_0^1 dx {f(x) \over (1-x)_+} = \int_0^1 dx {f(x) - f(1) \over 1-x}
\nonumber
\ee
for any smooth function $f(x)$.
Notice that the eqn.(\ref{asyq3}) is free of any soft singularity in 
the limit $m_g \rightarrow 0$.  The only divergent present in the above 
expression in this limit is collinear divergent.  This divergent
term is absorbed into the polarised anti-quark fragmentation function
and hence the fragmentation function is defined in the
massive gluon scheme.  Similarly, one can compute the gluonic
contribution when the quark is polarised.  We find this contribution
is same as that of anti-quark.   Comparing the 
eqn.(\ref{asyq3}) with eqn.(\ref{asym}) and including gluonic contribution
to polarised quark, 
we find  
\be
\hat g_1^q(x,Q^2)=3 {1 \over x} \sum_q e_q^2 \int_x^1 {dy \over y} 
{\cal C}_q(y,Q^2)
\left[ \Delta D_q^\Lambda(x/y,Q^2) + \Delta D_{\bar q}^\Lambda (x/y,Q^2)\right]
\label{g1q}
\ee
where 
\bea
{\cal C}_q(y,Q^2)& =& {4 \over 3}{\alpha_s \over 2 \pi} 
\left[\left({1+y^2 \over 1-y}\right)_+ \log\left({Q^2 \over \Lambda^2}\right) + 
(1+y^2) \left( {\log(1-y) \over 1-y}\right)_+
\right. \nonumber \\
&& \left. +\left ({1+y^2 \over 1-y}\right) \log(y) -  {3 \over 2} (1-y) -
{3 \over 2} \left({1 \over 1-y}\right)_+ \right. \nonumber \\
&& \left. - \left( {9 \over 4} - {\pi^2 \over 3} \right) \delta(1-y) \right]
\label{cq}
\eea

Now we turn to the computation of the cross section when the gluon
is polarised.  This amplitude is different from the amplitude
for polarised quark.  The differential cross section for the
production of polarised gluon can be formally written as
\be
\partg={i s \over 4 q_1.q_2} L_{\mu \nu} (q_1,q_2,s_1) {1 \over Q^4}
\epsilon_{\mu\nu\lambda\sigma} q^\lambda p_3^\sigma {Q \over p_3.q} H_g(p_3,q)
\label{asyg1}
\ee
where $x_3=p_3.q/Q^2$,$p_3$ is the momentum of the polarised gluon and $\theta$ is the
angle between produced polarised gluon and the incoming electron.  
The hard part of the
polarised gluon emission $H_g(p_3,q)$ is given by
\be
H_g(p_3,q)= {Q \over 32 (2 \pi)^3} \int dx_2 {\cal P}_g^{\mu \nu} \vert M^g 
\vert^{2}_{\mu \nu}
\label{hadg}
\ee
where the projector ${\cal P}_g^{\mu \nu} = 
i \epsilon_{\mu \nu \lambda \sigma} p_3^\lambda q^\sigma/2 p_3.q$.
The projected matrix element square ${\cal P}_g. \vert M^g\vert^2$ 
is computed from 
the Fig. 2(with gluon polarised) and is given by
\bea
{\cal P}_g^{\mu \nu} \vert M^g \vert^{2}_{\mu \nu}&=&{ 16 (2 \pi)^3 
e_q^2 \alpha \alpha_s \over  \pi} \left({m_g^2 Q^2-st \over (s+t)^2}\right) 
\left[ 2{(s+t)(s+t-m_g^2-Q^2) \over s t} \right. \nonumber \\
&& \left. +{ 4 m_g^2 Q^2 - 2 m_g^2 s - 2 Q^2 s +s^2 -t^2 \over t^2}
\right. \nonumber \\
&& \left. +{ 4 m_g^2 Q^2 - 2 m_g^2 t - 2 Q^2 t +t^2 -s^2 \over s^2}
 \right]
\label{prhadg}
\eea

Following the similar procedure adopted in the quark sector, 
we obtain the expression for
differential cross section for polarised gluon emission:
\be
\partG={4 e_q^2 \alpha^2 \alpha_s \over  Q^2} \left[(2-x_3) 
\log\left({1+ \beta_{x_3}\over 1- \beta_{x_3}} 
\right) - 2 (2-x_3) \beta_{x_3} \right] Cos(\theta)
\label{asyg2}
\ee
where $\beta_{x_3}=\left(1-4 m_g^2 / x_3^2 Q^2
\right)^{1/2}$.
The above result shows that there are no soft singularities.
The small nonzero gluon mass is used to regulate collinear
divergence.  This regulator dependent term is absorbed into the
polarised gluon fragmentation function at the level of full cross section,
hence, the polarised gluon fragmentation function is defined in the massive
gluon scheme.  Since we are looking at the polarised gluon production,
there is no virtual correction to this order, hence
there is no UV divergences.   Substituting the above equation 
in the parton model expression, we find the expression for $\hat g_1^g(x,Q^2)$ as
\be
\hat g_1^g(x,Q^2)=3 { 1 \over x} \sum_q e_q^2 \int_x^1 
{dy \over y} {\cal C}_g(y,Q^2)
\Delta D_g^\Lambda(x/y,Q^2) 
\label{g1g}
\ee
where the coefficient function ${\cal C}_g(y,Q^2)$ is given by
\be
{\cal C}_g(y,Q^2)=2 {4 \over 3} {\alpha_s \over 2 \pi} \left[  (2-y) 
\log({Q^2 y^2 \over \Lambda^2 }) + 2 (y-2)\right]
\label{cg}
\ee

In this paper we have discussed the importance of gluons in the
polarised $\Lambda$ production in $e^+ e^-$ scattering.
We have analysed this using the AP equation satisfied by the
quark and gluon fragmentation functions.  We have computed QCD
corrections to the fragmentaion function $\hat g_1^\Lambda(x,Q^2)$
to order $\alpha_s$ in the parton model.

It is a pleasure to thank Prof M.V.N. Murthy for his 
constant encouragement and fruitful discussions.
I would like to thank Prof H.S. Mani for his invitation 
to Metha Research Institute where this work was formulated
and I am also thankful to Prof H.S. Mani, Prof R. Ramachandran 
and Prof K. Sridhar for their carefull reading of the 
manuscript and valuable comments.
I thank Dr. Prakash Mathews for his valuable discussions.

\newpage
{\bf \Large Figure Captions:}
\begin{enumerate}
\item
Graph contributing to $e^-(q_1) e^+ (q_2)\rightarrow \bar q(p_1)  q(p_2)$.
\item
Graphs contributing to $e^-(q_1) e^+ (q_2)\rightarrow \bar q(p_1)  q(p_2)
g(p_3)$.
\item
Virtual corrections to the figure.1
\end{enumerate}
\end{document}